\documentclass[pre,twocolumn]{revtex4}

\usepackage{graphicx}
\usepackage{epsfig}
\usepackage{bm}
\usepackage{amsmath}
\usepackage{amssymb}
\usepackage{amsfonts}
\usepackage[normalem]{ulem}
\usepackage{subfigure}
\usepackage{fancyhdr}

\begin{document}

\title{Random M\"{o}bius Maps: \\ Distribution of Reflection in Non-Hermitian 1D Disordered Systems}
\author{Theodoros G. Tsironis}
\author{Aris L. Moustakas}
\affiliation{Department of Physics, National Kapodistrian University of Athens, Athens 15784, Greece.}


\begin{abstract}
Using the properties of random M\"{o}bius transformations, we investigate the statistical properties of the reflection coefficient in a random chain of lossy  scatterers. We explicitly determine the support of the distribution and the condition for coherent perfect absorption to be possible. We show that at its boundaries the distribution has Lifshits-like tails, which we evaluate.  We also obtain the extent of penetration of incoming waves into the medium via the Lyapunov exponent. Our results agree well when compared to numerical simulations in a specific random system.
\end{abstract}

\maketitle


\section{Introduction}

Spatial inhomogeneity and absorption are ubiquitous in all but very carefully manufactured transmission media, thereby often complicating their scattering properties. The relevance of disorder and coherent absorption in such systems has been extensively studied using numerical \cite{maschke1994electron, gupta1995electron} and analytic methods. Such methods include mainly random matrix theory (RMT) when the number of propagating modes is large \cite{Beenakker1997_MesoscopicReview,Li2017_RMTCPA, Fyodorov2017_DistributionZerosSmatrix_CPA} and, in the opposite limit, effectively one-dimensional models \cite{Gertsenshtein_Vasilev_1959, Kohler1976_Transmission_Line_Disorder_Loss, freilikher1994wave, pradhan1994localization,  Paasschens1996_Localization, heinrichs1997light, brouwer1998transmission}. In these models the individual scatterers are assumed weak and small compared to the wavelength, essentially forming a continuum $\delta$-correlated Gaussian potential. 

The recently discovered phenomenon of coherent perfect absorption (CPA) \cite{Chong2010_CoherentPA, Liu2010_CPA_PlasmonicSensor, Richoux2015_DisorderPeriodicArrayAcousticHelmholtzResonators, Romero2016_CPA_Acoustic} whereby an incident wave, which becomes localized in an absorbing region and is therefore strongly attenuated, can thus be traced directly to the interplay between inhomogeneity and absorption. While CPA is mostly studied in systems with a few carefully positioned absorbing scatterers, recent works have analyzed experimentally \cite{Pichler2019_CPA_ExperimentalRF} and numerically \cite{Liew2014_CPANumerical} the effects of randomness in the absorption characteristics of a larger system. It is therefore important to obtain a better understanding of the interplay between randomness and local absorption.  Towards this direction, in \cite{Li2017_RMTCPA} the authors applied random matrix theory to a random cavity with a few local absorbing elements to obtain the statistics of the modulus of the reflection coefficient. 

In this paper, we take a parallel approach and model a waveguide with a large number of discrete absorbing scatterers with properties from a specific experimental setup \cite{Romero2016_CPA_Acoustic}, spaced at random distances from each other. We assume that the waveguide is long compared to the  localization length. The one-dimensional geometry of the system we study can be justified by the fact that many realistic waveguides are often one-dimensional and recent experiments with lossy scattering elements are indeed one-dimensional \cite{Chong2010_CoherentPA, Richoux2015_DisorderPeriodicArrayAcousticHelmholtzResonators}. Hence it is also experimentally relevant to analyze one-dimensional waveguides with lossy scatterers. In contrast to the theoretical models discussed above, the distance between scatterers can be comparable to the wavelength. Nevertheless, an appropriate long-wavelength limit in our model can reduce to the  $\delta$-correlated Gaussian models. In the infinite-size limit we are considering, the total transmission probability through the sample is strictly zero. Therefore, in the absence of absorption, the reflection from the sample has unit norm, as expected in localized systems. When absorption is included, the resulting distribution of the reflection coefficient will display the competition between destructive interference, which will tend to coherently reflect the wave, and strong absorption in the regions, where the wave is in near resonance and undergoes many reflections. Unlike previous works, we deal with the distribution of the reflection on the complex plane, rather than just its modulus. We should point out that our approach here can be generalized to amplified disordered chains, providing information about the statistics of the \emph{amplification} \cite{pradhan1994localization,  Paasschens1996_Localization}, as well as to random waveguides with local parity-time reversal symmetry \cite{longhi2010_PTLaser}. 

The focus of the paper is threefold. First, we fully determine the support of the distribution of reflection in the complex unit disc. To be able to do this, we utilize the properties of M\"{o}bius maps, which are  conceptually simple but provide strong constraints on the support in the complex plane. Subsequently, we identify the Lifshits-like tails of the reflection distribution at the boundaries of its previously discussed support and analytically obtain the asymptotic behavior of the distribution. Finally, we establish the relationship connecting the reflection distribution with the penetration length of the system. Such a relationship is known to exist, since the reflection distribution is known to contain all the information regarding the localization properties of the system \cite{Anderson1980_ScalingLocalization}.

\section{Iterative Random M\"{o}bius Maps}
\begin{figure}
    \centering
    \includegraphics[width=.95\columnwidth]{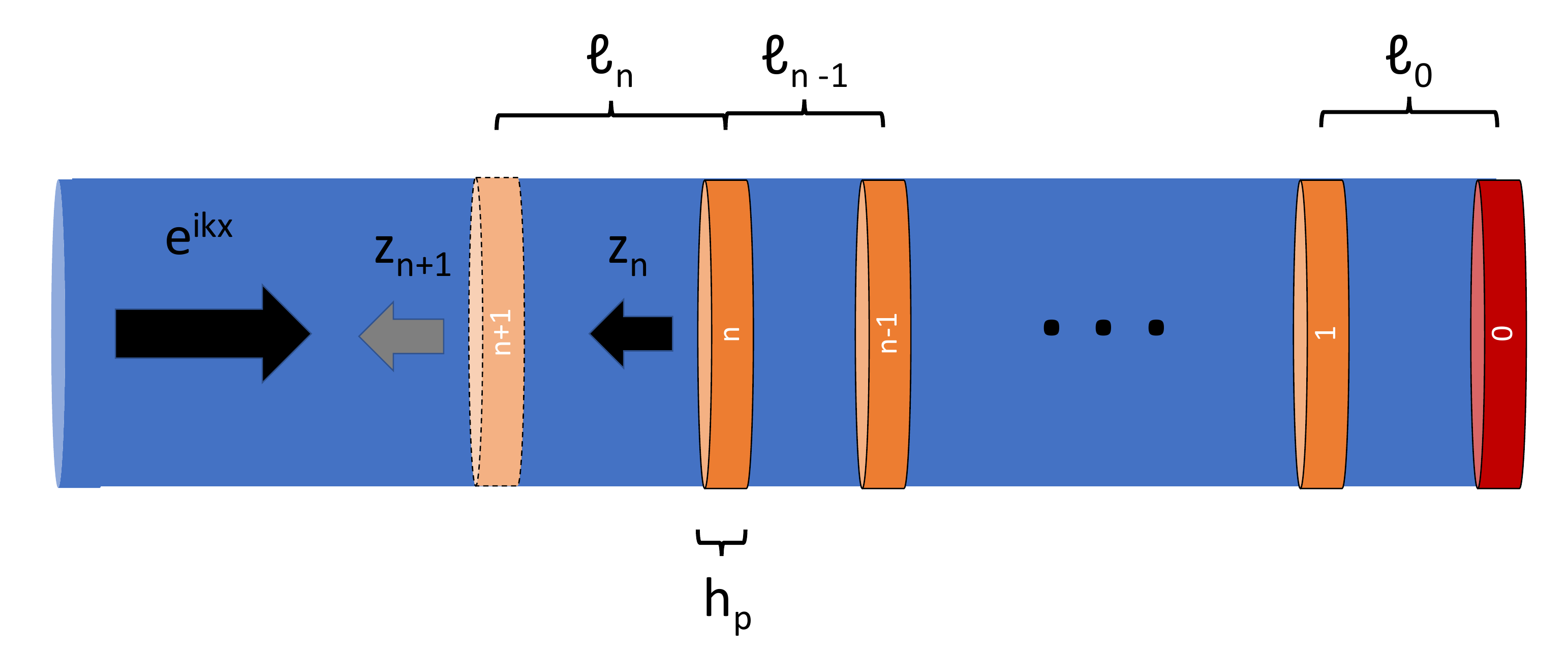}
    \caption{The construction of the system is depicted. We place the $n+1$-th scatterer at distance $\ell_n$ from the system, between the system and the beam, changing its reflection from $z_n$ to $z_{n+1}$. In later figures we will provide data for an element characterised by its thickness $h_p$. The terminating element, depicted in red, will not matter at the large system limit.}
    \label{fig:Geometry}
\end{figure}

The purpose of this paper is to analyze the reflection properties of a one-dimensional half-infinite waveguide consisting of absorbing scatterers located at random distances from each other. In order to simplify the following notation, we will work with parity-symmetric scatterers; however, the generalization to parity breaking scatterers is trivial. To obtain the total reflection coefficient we start with a single scatterer with scattering matrix
\begin{align}
    S=\left[\begin{array}{cc}
        \rho & \tau \\
         \tau & \rho
    \end{array}
    \right]
\end{align}
with $\rho$ and $\tau$ being the reflection and transmission coefficients, respectively,  at a given wavenumber $k=2\pi/\lambda$ corresponding to wavelength $\lambda$.  We assume that the scatterer is lossy (and hence non-unitary), with the absorbed power fraction given by $1 - |\tau|^2 - |\rho|^2>0$. Then, we iteratively place an additional scatterer at a random distance $\ell$ to the left of the last one, see Fig. \ref{fig:Geometry}. We may then express $z_{n+1}$, the total reflection coefficient after placing the ($n+1$)th scattering element, in terms of the corresponding reflection coefficient $z_n$  and the scattering data of the last scatterer as follows \cite{Beckenbach2013_Mathematics_for_the_Engineer_book}
\begin{align}
\label{Combination_of_Obstacles}
z_{n+1}=\rho+\frac{ \tau^2 z_n e^{i\phi_n}}{1- \rho z_n e^{i\phi_n}} \equiv M\left(z_n e^{i\phi_n}\right),
\end{align} 
where we suppress the dependence of the scattering data from $k$ and 
$\phi_n = 2k\ell_n$, in which $\ell_n$ is the random distance between scatterers $n$ and $n+1$. The above expression follows from combining the scatttering properties of two entities, namely the $n$th scatterer and the one-dimensional waveguide segment of length $\ell_n$. We assume that the distances between scatterers are identically and independently distributed. For simplicity, we also assume that the phase distribution $\mu(\phi)$ does not vanish anywhere. 
Notably, the reflection coefficient at the $n$th step, depends on the system to the right only through its reflection coefficient $z_{n-1}$, thereby making the above iterative process a Markov chain. The initial condition $z_0$ depends on the termination process on the right. For an open system we have $z_0=0$, while for a fully reflecting termination we have $|z_0|=1$. Our focus will be to understand the behavior of the distribution of reflection coefficients in the limit $n\to\infty$, for which the initial condition will eventually no longer matter. Indeed, this is the case here for two reasons. First, because the wave will have been attenuated by the time it reaches scatterers far from its incidence. In addition, due to the one-dimensional geometry of the waveguide, randomness in the position of the scatterers will eventually make sure that the transmission coefficient will become negligible. 

It is worth mentioning here that the above map can also provide the behavior of a random laser by time-reversing the parameters using the time-reversal operation of the scattering matrix of the scatterer $S\to (S^*)^{-1}$, i.e. $\rho\to \tilde{r}/d^*$, $r\to \rho^*/d^*$ and $d\to 1/d^*$, where $S$ is the total system scattering matrix, while $d$ is the determinant of the single-scatterer scattering matrix, i.e. $d=\rho^2-\tau^2$. Then since the transmission coefficients of the total system will remain zero due to localization, time reversal maps the zeros of the original absorbing system to poles of the resulting time-reversed amplifying system in the sense that 
$z\to v=1/z^*$ in the sense that
\begin{align}
\label{TR_recursion}
v_{n+1}=\frac{\frac{\rho^*}{d^*}-\frac{1}{d^*} v_n e^{-i\phi_n}}{1- \frac{\tilde{r}}{d^*} v_n e^{-i\phi_n}}
\end{align}

The function $M(z)$ is a general M\"{o}bius transformation and is therefore  conformal in $\mathbb{C} \setminus \{ \rho^{-1} \}$ and maps circles onto circles \cite{Jones1987_Moebius_Geometry_book}. This is a key property for our random sequence. Observe in \eqref{Combination_of_Obstacles} that $z_{n+1}$ is constructed from $z_n$ after a random rotation about zero and then an action of $M$. Hence, the possible values of $z_{n+1}$ for a fixed $z_n$ lie on a circle. Specifically, $M\left(x e^{i\theta}\right)$ for $\theta\in[0,2\pi)$ and $0<x\leq 1$ maps the circle of radius $x$ around the origin to a circle \cite{Beckenbach2013_Mathematics_for_the_Engineer_book} centered at 
\begin{align}
    c_0(x) = \rho + \frac{\tau^2 \rho^*x^2}{1-|\rho|^2 x^2}
\end{align}
with radius 
\begin{align}
r_0(x) = \frac{x|\tau|^2}{1-|\rho|^2 x^2}.    
\end{align}
The maximum and minimum distances from the origin are then given by the functions $F_+$ and $|F_-|$, where 
$F_\pm(x) = \left| c_0(x)\right| \pm r_0(x)$. 

We restrict this study to reflectively dissipative systems, for which $F_+(1) \leqslant 1$. This condition is always satisfied by parity-symmetric lossy scatterers as they have been described above. Such iterative processes are commonly described \cite{Lifshits1988_TheoryDisorderedSystems_book} by the evolution of the probability distribution $P_n$ of the reflection amplitude, as a result of the action of an operator $\cal{M}$ corresponding to the Markov process described above, on the initial distribution $P_0$.
${\cal M}$ is defined as
\begin{eqnarray}
\label{Markov_Evolution}
&& {\cal M}P(re^{i\theta}) = \iint dr' d\theta' P(r'e^{i\theta}) \int d\phi \mu(\phi) \nonumber \\
&& \delta(r - |M(r' e^{i(\theta'+\phi)})|) \delta(\theta - \arg( M(r' e^{i(\theta'+\phi)}))
\end{eqnarray}
The sequence $P_n = {\cal M}^nP_0$ is compactly supported within the unit circle and since $\cal M$ is continuous in the weak topology, the limit $P = \lim_{n\to\infty} {\cal M}^n P_0$ is well defined and is a stationary point of ${\cal M}$. 

\section{Support of Steady-State Distribution}

To obtain the support of the limiting distribution, we first note that 
when $\rho \neq 0$, $F_+(r) = r$ has at least one root in $[0,1]$, the smallest of which can also be obtained as the limit $\tilde{r} = \lim_n F^n_+ (0)$. Correspondingly, there exist unique angles $\tilde{\theta}$, $\tilde{\phi}$, such that $\tilde{z}=\tilde{r}e^{i\tilde{\theta}}$ and $\tilde{z}=M\left(\tilde{z} e^{i\tilde{\phi}}\right)$. Typically, this point is different than the non-random fixed point $\bar{z}=M(\bar{z})$.

We consider the disk enclosed within the circle $M( \tilde{r} e^{i\phi})$, with $\phi\in(0,2\pi]$, which we represent as $M(D(\tilde{r}))$. If this disk includes the origin, i.e. if $F_-(\tilde{r})<0$, then we shall show that the support of the steady state distribution $P(R,\theta)$, is simply $M(D(\tilde{r}))$. On the other hand, if the origin is not included in this disk, i.e. $F_- (\tilde{r})>0$, then the support of $P(\cdot)$ is the $M(D(\tilde{r}))\setminus M(D(F_-(\tilde{r})))$, i.e the disk $M(D(\tilde{r}))$ after excluding the  image, under $M(\cdot)$, of the disk around the origin with radius $F_-(\tilde{r})$.

To prove the above assertion for the support,  we start by noting that $P(\cdot)$ has no support for $|z|>\tilde{r}$, because in this region we have $F_+(r)<r$, and hence under iteration this region is depleted.
Within the disk $D(\tilde{r})$, the probability mass also has to reside within the disk $M(D(\tilde{r}))$. If not, i.e. if there were a finite probability to be within $D(\tilde{r})$, but not in $M(D(\tilde{r}))$, this would mean that at a previous iteration this region would have been outside $D(\tilde{r})$. Additionally, if $F_-(\tilde{r})>0$, the disk $D(F_-(\tilde{r}))$ is outside $M(D(\tilde{r}))$. Hence its image under $M(\cdot)$ cannot be in the support of $P(\cdot)$. As a result, in this case, the support is included in $M(D(\tilde{r}))\setminus M(D(F_-(\tilde{r})))$. 

Starting with the case  $F_-(\tilde{r})\geq 0$, it suffices to show that any point within $M(D(\tilde{r}))\setminus M(D(F_-(\tilde{r})))$ is accessible from any other point in the same domain. Starting with any point $z$ in $M(D(\tilde{r}))$, with radius $0\leq |z| <\tilde{r}$ and any small number $\epsilon>0$. Then, since $\lim_{n\to\infty}F_+^n(|z|)= \tilde{r}$, there is finite probability that after a finite number of steps $m$ its image $z_m$ will be  $|z_m-\tilde{z}|<\epsilon$. Therefore, after the next random M\"{o}bius transformation, with finite probability $z_{m+1}$ will have radius close to any radius in $(F_-(\tilde{r}),\tilde{r})$. As a result, after the final transformation, $z_{m+2}$ can be close to any point in $M(D(\tilde{r}))\setminus M(D(F_-(\tilde{r})))$.

In the case where $F_-(\tilde{r}) < 0$, the above argument can be directly applied to show that $M(D(\tilde{r}))\setminus M(D(|F_{-}(\tilde{r})|))$ is within the support. To show that $M(D(|F_{-}(\tilde{r})|))$ is also included in the support, it is sufficient that any point within $D(|F_{-}(\tilde{r})|)$ can be reached.
Starting from any point with radius in $(|M^{-1}(0)|,\tilde{r})$, since for points with such radii $F_{-}(r) < r$, there is a finite probability to move to lower radii. Since also $F_{+}(r) > r$, at some point the radius $|M^{-1}(0)|$ will be reached, from where any range in $(0,F^{+}(|M^{-1}(0)|)$ can be reached and hence any point in $D(|F_{-}(\tilde{r})|)$ is in the support.


From the above analysis, it follows that the condition for CPA to be possible is simply 
\begin{align}\label{eq:CPA_condition}
F_-(\tilde{r})=\left| c_0(\tilde{r})\right|-r_0(\tilde{r}) <0,    
\end{align}
a condition, which can be shown to be equivalent to $|\rho| < \tilde{r} |d|$. 

One surprising result of the above analysis is that as long as $\mu(\phi)$ has full support, the condition for CPA to be possible in \eqref{eq:CPA_condition} is independent of the \emph{details} of $\mu(\phi)$, but depends only on the scatterers' properties. We can see this from the iterative expression of the reflection coefficient in \eqref{Combination_of_Obstacles}. For the total reflection after $z_N$ (where $N$ is the total  number of scatterers) to be zero, one needs the modulus of the reflection coefficient excluding the final scatterer to be $|z_{N-1}|=|\rho|/|\rho^2-\tau^2|$. However, as we have  argued above, for sufficiently large $N$ the scattering amplitude is within a circle of radius $\tilde{r}$, hence  $|z_{N-1}|<\tilde{r}$. Therefore, only if $|\rho|>|\rho^2-\tau^2|\tilde{r}$, in the case that $\mu(\phi)$ has full support, is there a finite probability that a number of scatterers close to the end of the transmission line are spaced appropriately, so that the the condition $z_N=0$ is met. Thus, the form of $\mu(\phi)$ will only determine how large the probability for CPA will be. On the other hand, if the support of $\mu(\phi)$ has gaps, then there exist scatterers, for which the above ``perfect arrangement'' has zero probability, thus making CPA impossible. However, for this to happen the zero-support of $\mu(\phi)$, which is $k$-dependent, will have to match that of the properties of the scatterer. As a result, this is a very scatterer-dependent issue, which we will not discuss further.

\begin{figure}
    \centering
    \includegraphics[scale=.38]{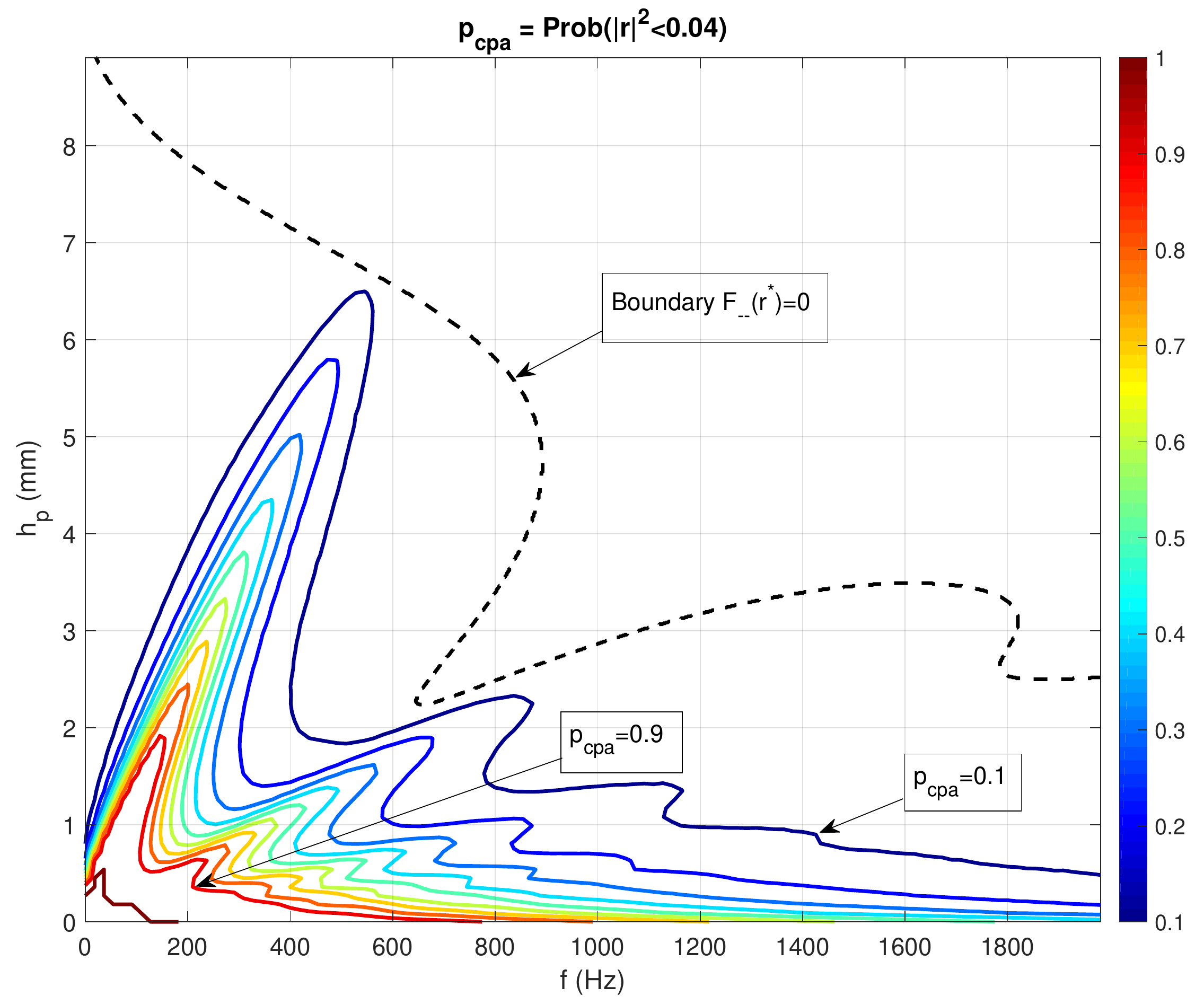}    
    \caption{Probability over $2\cdot 10^5$ realizations for the reflection power to be close to zero ($P(|r|^2<0.04)$) in the case of a  large array of Poisson-distributed spacings (mean $\ell_0$=20cm) between lossy, acoustically resonant plates (see \cite{Romero2016_CPA_Acoustic}). The y-axis of the plot is the thickness of the lossy plates, which is a measure of their lossy behavior. The colored curves represent the contours of constant probability, from $p=0.1$ to $p=0.9$. The dashed curve represents the boundary, to the right of which the probability density vanishes at $r=0$. The probability is typically, but not always, a decreasing function of absorption, parametrized by $h_p$.} 
    \label{fig:Piezoelectric_Existence_of_CPA}
\end{figure}

\begin{figure}
    \centering
    \includegraphics[scale=.5]{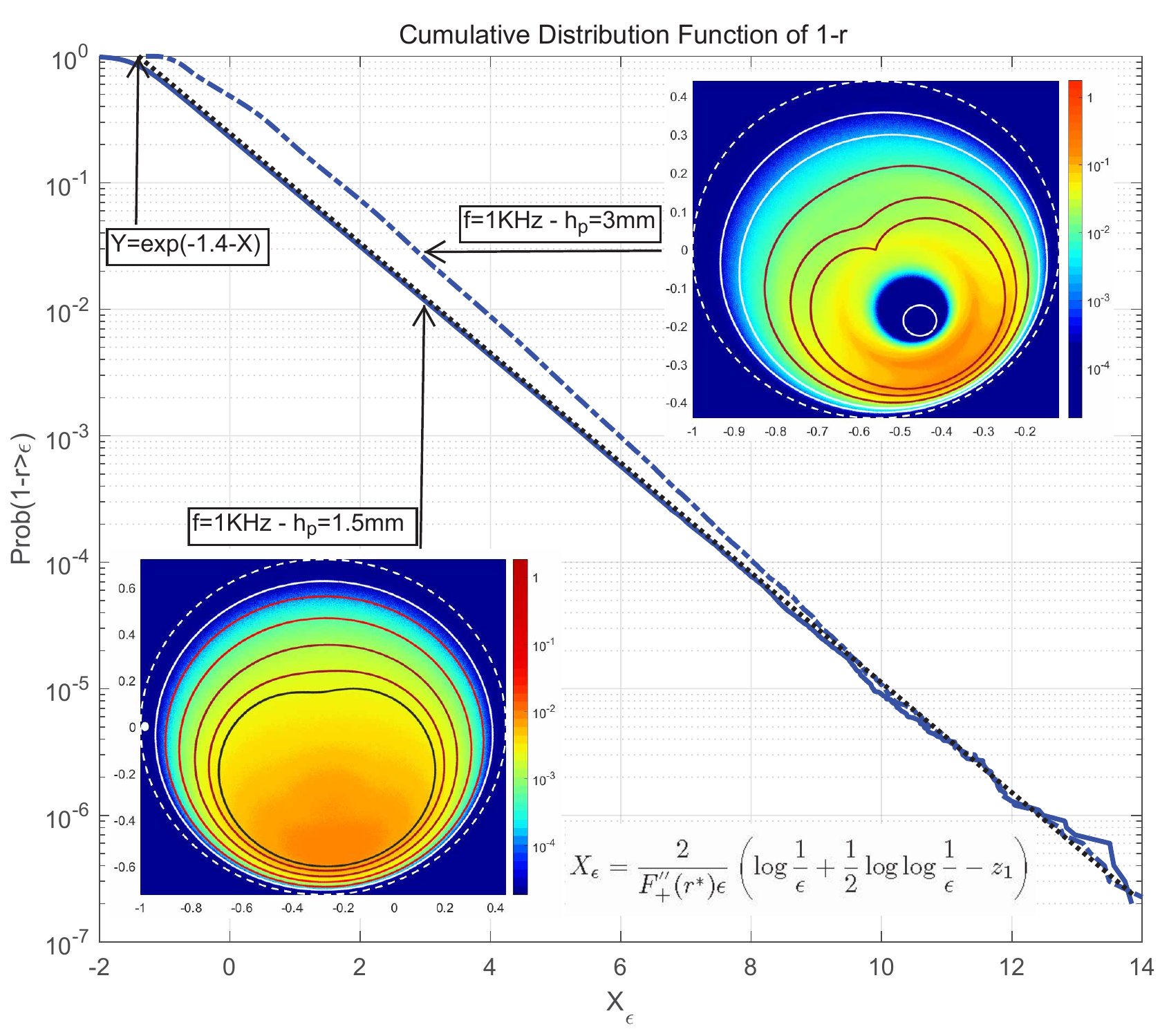}
    \caption{Cumulative distribution function (CDF) $\Pr(1-|r|>\epsilon)$ for the tails of the distribution $p(\epsilon)$ in the case of a large array of randomly spaced lossy, acoustically resonant plates, at frequency $f=$1kHz and plate thickness $h_p$, as in Fig. \ref{fig:Piezoelectric_Existence_of_CPA}.  The CDF is plotted versus the exponent $X_\epsilon$ appearing  in the rhs of \eqref{p_gamma}. The insets depict the color-shaded distribution of $r$ over its support, in which the white dashed lines represent the boundaries of the support of the distribution, while the solid lines depict level contours.}
    \label{fig:Piezoelectric_Tails}
\end{figure}

\section{Tails of the Limiting Distribution}
The limiting distribution $P(z)$ satisfies the integral equation $P={\cal M}P$. Despite its complexity, we will use the properties of M\"{o}bius transformations to explore its asymptotic behavior close to the edges of its support. As we shall see, the distribution has Lishits-like tails \cite{Lifshits1988_TheoryDisorderedSystems_book} at the edges, which cannot be easily obtained by numerical calculations. 

To obtain the tails of the distribution, we start with the behavior near the point $z=\tilde{z}$, which will determine the asymptotics close to the entire boundary of $M(D(\tilde{r}))$. Accordingly we denote $p(\epsilon , \theta)=P((\tilde{r}-\epsilon)e^{i\theta})$ and $p(\epsilon) = \int d\theta p(\epsilon,\theta)$. Integrating $P = \mathcal{M}P$ over $\theta$ we get
\begin{eqnarray}
\label{p_r_Inteq}
p(\epsilon) = \iint d\epsilon' d\theta' p(\epsilon',\theta') \int d\phi \mu(\phi) \frac{\tilde{r}-\epsilon'}{\tilde{r}-\epsilon}\\ \nonumber
\delta(\tilde{r} - \epsilon - |M((\tilde{r} - \epsilon') e^{i(\phi+\theta')})|).
\end{eqnarray}
The argument of the $\delta$-function vanishes when the angle $\phi+\theta'$ is such that the image of the circle with radius $\tilde{r}-\epsilon'$ intersects the circle with radius $\tilde{r}-\epsilon$. We define the inverse function $\phi+\theta'=\Theta(\epsilon,\epsilon')$. As a result we have
\begin{eqnarray}
\label{p_r_Inteq2}
p(\epsilon) = 2\iint d\epsilon' d\theta' p(\epsilon',\theta') \mu(\Theta(\epsilon',\epsilon)-\theta') \frac{\partial \Theta(\epsilon,\epsilon')}{\partial \epsilon},
\end{eqnarray}
where the factor of $2$ appears due to symmetry. Now for small $\epsilon$, the range of $\theta'$ is $\theta'= \tilde{\theta}\pm \Delta(\epsilon)$, where $\Delta(\epsilon)=\sqrt{\frac{2\epsilon}{\tilde{r}+|\tau|^2/(1-|\rho|^2\tilde{r}^2)}}$.  For small $\epsilon$, we may therefore approximate the argument of the function $\mu(\phi)$  in the integral with $\tilde{\phi}$ and simplify the above expression to 
\begin{eqnarray}
\label{p_r_Inteq3}
p(\epsilon) \approx 2\mu(\tilde{\phi})\int d\epsilon' p(\epsilon')  \frac{\partial \Theta(\epsilon,\epsilon')}{\partial \epsilon}.
\end{eqnarray}Additionally, for small $\epsilon$, the integration range is $\epsilon'\in(0,\epsilon_{\max})$, where $\epsilon_{\max}$ is the solution of the equation $\tilde{r}-\epsilon=F_+(\tilde{r}-\epsilon_{\max})$, which can be written to leading order as
\begin{eqnarray}
\label{eps_max}
\epsilon_{\max}=\frac{1}{F'_+(\tilde{r})}\epsilon+\frac{F^{''}_+(\tilde{r})}{2\left(F^{'}_+(\tilde{r})\right)^3}\epsilon^2.
\end{eqnarray}
It is easy to see that $F'_+(\tilde{r})\leq 1$ and that if $F'_+(\tilde{r})= 1$ then $F{''}_+(\tilde{r})>0$. Indeed, since for $x\leq \tilde{r}$, we have $F_+(x)\geq x$, we take $x=\tilde{r}-\epsilon$ with $\epsilon$ small. Then expanding around $\epsilon=0$ we get $\tilde{r} > F_+(\tilde{r})-F'_+(\tilde{r})\epsilon+F^{''}_+(\tilde{r})\epsilon^2/2>\tilde{r}-\epsilon$, which gives the required result.

Approximating the derivative in \eqref{p_r_Inteq3} close to $\epsilon=\epsilon'=0$ we obtain
\begin{eqnarray}
\label{p_r_Inteq4}
p(\epsilon) = C\int_0^{\epsilon_{\max}}  d\epsilon' \frac{p(\epsilon')}{\sqrt{\epsilon_{\max}-\epsilon'}},
\end{eqnarray}
where
\begin{align}
\label{C_expression}
C=\frac{\sqrt{2}\mu(\tilde{\phi})\left|\rho-(\rho^2-\tau^2)r^{*}\right| }{(1-|\rho|^2 \tilde{r}^2)\sqrt{r^{*3}|c_0(\tilde{r})|r_0(\tilde{r})F'_+(\tilde{r})}}.
\end{align}

Expressing the probability distribution as $p(\epsilon) = e^{-I(-\log(\epsilon))}$, where $I(t)$ an unknown function, and inserting this in (\ref{p_r_Inteq4}) and integrating the right-hand side asymptotically around $\epsilon'=\epsilon_{\max}$ we obtain the relation
\begin{align}
\label{p_r_Inteq7}
I(\log(\frac{1}{\epsilon_{\max}}))-I(\log(\frac{1}{\epsilon})) = \frac{1}{2}\log\frac{C^2\pi\epsilon_{\max}}{4I'(\log(\frac{1}{\epsilon_{\max}}))}. 
\end{align}
From this expression we may deduce the asymptotic behavior of $I(t)$ as $t\to\infty$. There are two cases we need to distinguish. 

Specifically, when $F'_+(\tilde{r})<1$ the solution the distribution $p(\epsilon)$ is essentially lognormal with logarithmic corrections
\begin{align}
\label{p_lognormal}
-\log p(\epsilon) &= \frac{\left(\log\epsilon\right)^2-2\log\epsilon\log\log\frac{1}{\epsilon}+2q_0\log\epsilon}{4\log \frac{1}{F'_+(\tilde{r})}},
\end{align}
with $\mathcal{O}(1)$ corrections and
\begin{align}
\label{z0}
 e^{q_0}=\frac{C^2\pi e\log\frac{1}{F'_+(\tilde{r})}}{2\sqrt{F'_+(\tilde{r})}}.
\end{align}

In contrast, when $F^{'}_+(\tilde{r})=1$, the distribution falls much faster to zero close to $r=\tilde{r}$. In this case
\begin{align}
\label{p_gamma}
-\log p(\epsilon) &= \frac{2}{F^{''}_+(\tilde{r})\epsilon}\left(\log\frac{1}{\epsilon}+\frac{1}{2}\log\log\frac{1}{\epsilon}-q_1\right),
\end{align}
with correction of order $o(1)/\epsilon$ and
\begin{align}
\label{z1}
e^{q_1}=Ce\sqrt{\frac{F^{''}_+(\tilde{r})\pi}{2}}.
\end{align}

We will now obtain the behavior close to the outer borders of the support i.e. at location $R(\delta,\chi)=c_0(\tilde{r})+(r_0(\tilde{r})-\delta)e^{i\chi}$. We observe that any such point $z$ originates, before a previous application of the random  M\"{o}bius map, from a point $z'$ in the neighborhood of $\tilde{z}$, in the sense that $z=M\left(z'e^{i\phi}\right)$, where $\phi\in(0,2\pi]$. Therefore, we can use the results we got above to obtain, to leading exponential order,
\begin{align}
\label{p_boundaries1}
P(R(\delta, \chi)) &\approx  p(\epsilon'(\delta,\chi)) \mu(g(|R(\delta, \chi)|,\tilde{r})- \tilde{\theta}) \\ \nonumber
& \frac{\partial  g}{\partial r}(|R|,\tilde{r})
\frac{\partial  \bar{R}}{\partial \theta}(|R|,\arg R),
\end{align}
where
\begin{align}\label{epsilon'(delta,chi)}
\epsilon'(\delta,\chi)) = \frac{(1-|\rho|^2 \tilde{r}^{2})^2}{|1+|\rho| \tilde{r}e^{i(\chi-\tilde{\phi})}|^2} \frac{\delta}{|\tau|^2}
\end{align}
and we have defined $\bar{R}(r,\theta)$ and $g(r,r')$ through $\bar{R}=|M^{-1}(re^{i\theta})|$ and $r=|M(r'e^{ig})|$. The product of their partial derivatives appearing in (\ref{p_boundaries1}) can be expressed as
\begin{align}
& \frac{\partial  g(|R(\delta, \chi)|,\tilde{r})}{\partial r}
\frac{\partial  \bar{R}(|R(\delta, \chi)|,\arg R(\delta, \chi))}{\partial \theta} \nonumber \\ 
&= \frac{|r-\tilde{r}e^{i\chi}|^2\left(r^2|d-\rho \tilde{r}e^{i\theta}|^2-d^2|r-\tilde{r}e^{i\theta}|^2\right)}{\tilde{r}|\tau|^4\left(d^2-\rho^2 \tilde{r}^{2}\right)}.
\end{align}

Similarly, we can calculate the vanishing behavior of the probability distribution at the boundaries of $M(D(|F_{-}(\tilde{r})|))$, if it exists. 

In Fig. \ref{fig:Piezoelectric_Tails} we plot the tails of the cumulative distribution function $\Pr(1-|r|>\epsilon)$ for a large array of lossy, acoustically resonant plates  discussed in \cite{Romero2016_CPA_Acoustic, Bongard2010_AcousticTransmission} and separated at uniformly random distances from each other, with maximum distance $\pi/k$. We have chosen two different values of plate thickness, which quantifies lossiness, for which the distribution has and has not, respectively, finite probability of coherent perfect absorption. The tails agree with the theoretical curve to a significant degree with the theoretical prediction of the tails in \eqref{p_gamma}. The insets depict the color-shaded distribution of $r$ over its support, in which the white dashed lines represent the boundaries of the support of the distribution, while the solid lines depict level contours. These can be seen to follow the color variations and hence are relatively close to the actual level sets.

\begin{figure}
    \centering
    \includegraphics[scale=.475]{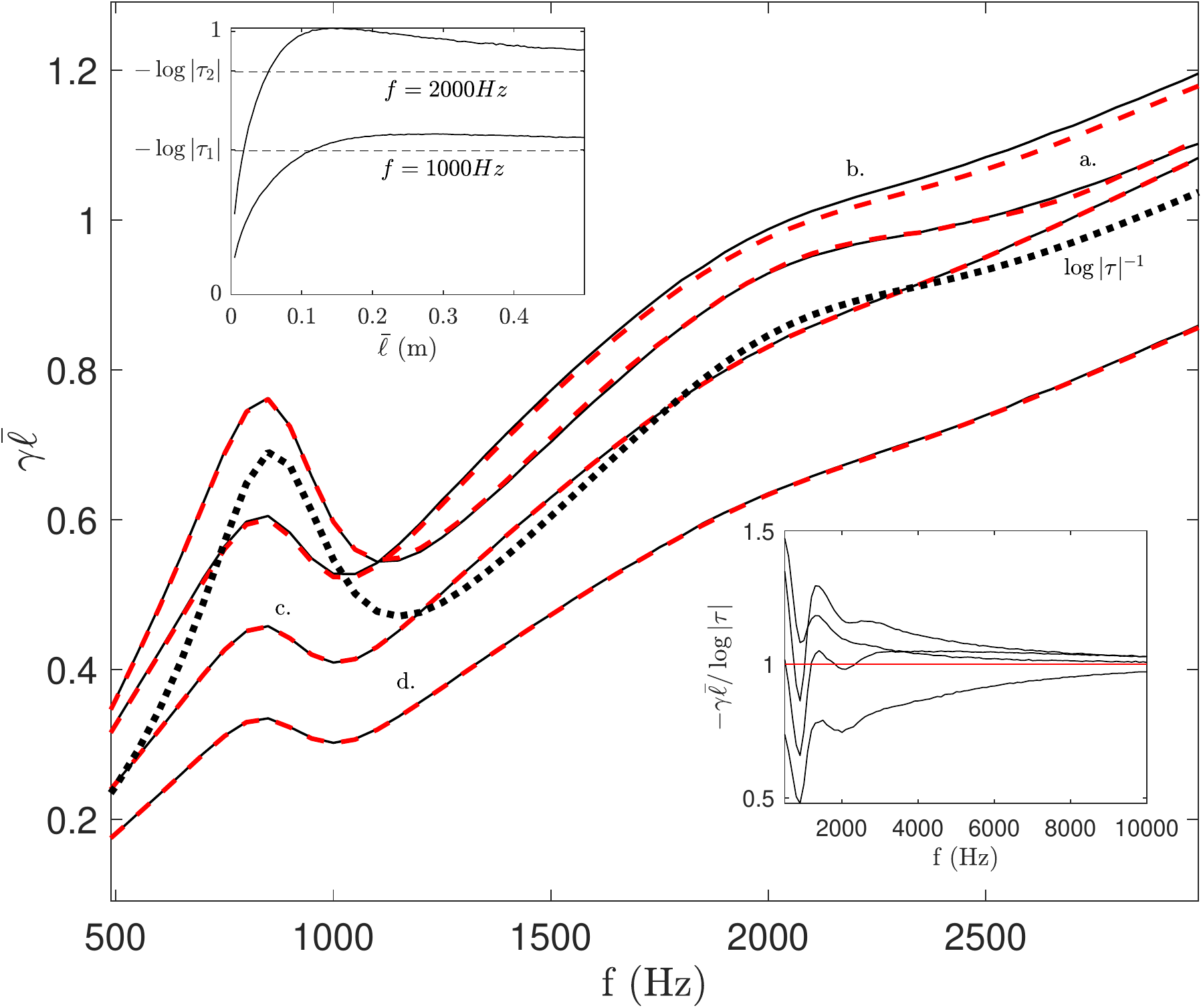}
    \caption{The normalized Lyapunov exponent $\gamma\bar{\ell}$ as a function of frequency, in case of lossy acoustic resonant plates that are Poisson-distributed  with average distance $\bar{\ell}$ a. $\bar{\ell} = 50 cm$ b. $\bar{\ell} = 10 cm$ c. $\bar{\ell} = 5 cm$ d. $\bar{\ell} = 2.5 cm$.  The solid lines correspond to the values obtained from \eqref{Lyapunov_vs_Reflection} by numerically generating the distribution of reflection and integrating over $\mu(\phi)$. The colored dashed lines were generated from \eqref{Lyapunov}, by calculating $|A_{n-1}|/|A_n|$ from \eqref{Lyapunov_Sequence}  with values of $z_n$ for a single instantiation of distances \emph{$\ell_n$} between $10^5$ scatterers. We see that the two sets of curves are practically identical. The bottom subplot shows the convergence of $-\gamma\bar{\ell}/\log|\tau|\to 1$ at high frequencies. On the top left subgraph we plot $\gamma \bar{\ell}$ vs. $\bar{\ell}$ for two frequencies, demonstrating the convergence to $-\log |\tau|$.}
    \label{fig:Piezoelectric_Lyapunov_vs_Wavenumber}
\end{figure}

\section{Penetration Length}
The distribution of reflection coefficients $P(z)$ can be used to obtain the penetration length of a wave with wavenumber $k$ incoming to the disordered medium. This is the so-called Lyapunov exponent of the system \cite{Lifshits1988_TheoryDisorderedSystems_book}, expressed as 
\begin{align} 
\gamma = -\lim_{x \to \infty} \log|\Psi(x)|/x,
\end{align}
where $\Psi(x)$ the corresponding wave function. 
For simplicity, in the next equations, we count the scatterers starting from the far left of the structure, rather than the far right, following the convention $m=N-n$, thus indexing the furthest left scatterer with $m=0$ ($n=N$), the one to its right $m=1$, etc.
Thus $z_m$ is now the reflection of the system if we remove the leftmost $m$ scatterers, $x_m$ is the location of the $m$-th scatterer $\ell_m = x_{m+1} - x_m=x_{N-n+1}-x_{N-n}$, etc. Now, for $x\in(x_m,x_{m+1})$ we may express the wave as a sum of left-moving and right-moving parts
\begin{align}
\Psi(x) = A_m e^{ik(x-x_m)} + B_m e^{-ik(x-x_m)},    
\end{align} 
where the second coefficient $B_m$ can be related to the first, $A_m$, through the reflection coefficient to the left of the $m+1$th scatterer as $B_m=z_m e^{2ik\ell_m} A_m$. We may now relate $A_m$ with $A_{m-1}$ by using the input-output relation of the scattering matrix at the position of the $m$-th scatterer
\begin{align} 
\label{Lyapunov_Sequence}
A_{m} = e^{ik\ell_{m-1}}\tau A_{m-1} + e^{2ik\ell_{m}}\rho z_{m} A_{m}. 
\end{align}
which expresses the fact that the wave travelling further inside in the system after the $m$-th scatterer must be the sum of the wave transmitted through that scatterer from the outer system towards the inside and what was reflected from the scatterer back inside the system. We may then express the Lyapunov exponent as
\begin{align}
\label{Lyapunov}
\gamma = - \lim_{N \to \infty} \frac{1}{N\bar{\ell}} \sum_{m=1}^N \log \left|\frac{A_{m}}{A_{m-1}}\right|.
\end{align}
Since $\ell_{m}$ is independent from $z_m$, we may express \eqref{Lyapunov} as a disorder average \cite{Lifshits1988_TheoryDisorderedSystems_book}, resulting in
\begin{equation}
\label{Lyapunov_vs_Reflection}
\gamma\bar{\ell} = \int \int r dr d\theta P(re^{i\theta}) \int d\phi \mu(\phi) \log \left| \frac{1-e^{i(\phi+\theta)}\rho r}{\tau} \right|,
\end{equation}
where the phase $\phi=2k\ell$ and $\ell$ is the inter-scatterer distance. 
When the distribution $\mu(\phi)$ is uniform over $\phi\in[0,2\pi)$, then the Lyapunov exponent becomes simply $\gamma\bar{\ell}=-\log|\tau|$. This is the case when, for example, the inter-scatterer distance is uniform over an integer multiple of $\pi / k$, or if it is Poisson-distributed with $k\bar{\ell}\gg 1$. In the opposite limit of the latter case, i.e. when  $k\bar{\ell}\ll 1$, the distribution $\mu(\phi)$ will be localized around 0, and thus $P(\cdot)$ will be localized around the fixed point of $M$, where $\bar{z}=M(\bar{z})$. As a result, $\gamma\bar{\ell}\approx-\log|\tau/(1-\rho \bar{z})|+O(k\bar{\ell})$.

In Fig.~\ref{fig:Piezoelectric_Lyapunov_vs_Wavenumber} we compare the penetration length for resonant plates as a function of frequency for various incident frequencies as the integral in (\ref{Lyapunov_vs_Reflection}) calculated by Monte Carlo methods and as the limit (\ref{Lyapunov}) calculated for a fixed instantiation of the random configuration. The agreement demonstrates the ergodicity of the process. 

\section{Discussion}
We have studied the distribution of reflection in a one-dimensional half-infinite line of randomly spaced lossy scatterers, taking  advantage of the properties of random Moebius transformations. We have found that the support of the distribution, and have shown that the scatterer (and not the spacing) properties solely determine the possibility of the occurrence of  coherent perfect absorption (CPA), determining the analytic condition for this to happen, when the scatterer-spacing distribution is such that the angle $\phi=2k\ell\mod(2\pi)$ has full support. 

We have only focused on the case where the distribution $\mu(\phi)$ has full support, i.e. when there are no regions of forbidden distances between the scatterers. If such regions are allowed to exist, then the M\"{o}bius map will map arcs to arcs instead of circles to circles. As a result, the condition for CPA to be possible will decrease and different asymptotic behaviors near the edges of the spectrum may appear. 

In a similar fashion, when the scatterer properties themselves are random, the criterion for CPA appearance discussed above is generalized to 
\begin{align}
\label{random_scatterers}
\text{Prob}(|\rho|<|\rho^2-\tau^2|\tilde{r})>0
\end{align} 
where the probability is over the distribution of scatterers. In this context, it is not surprising that in random $\delta$-correlated models discussed in the introduction, for which the scatterers are random, CPA is typically possible \cite{Kohler1976_Transmission_Line_Disorder_Loss,  pradhan1994localization}. Beyond $\delta$-correlated potentials, in numerical studies such as \cite{gupta1995electron} it can be seen also be seen that the possibility of CPA is dependant on the properties of the allowed scatterers.

We have also calculated the probability of CPA as functions of frequency and absorption, for a specific waveguide realization with lossy acoustic resonant plates \cite{Romero2016_CPA_Acoustic}.
In addition, we have found that special points on the circle of the outer boundary of support of the reflection determine the Lifshitz-like tails at all of its boundaries, which we have determined analytically. 
Finally, we have evaluated the Lyapunov exponent, which characterizes the extent of penetration of incoming waves into the system and discussed its properties. 
The above analysis can be extended to randomly placed scattering elements with amplification using the mapping \eqref{TR_recursion}. 

\section*{Acknowledgment} We would like to  thank G. Theocharis for interesting discussions. This work was supported in part by ONR Global Grant N62909-18-1-2141.
\bibliographystyle{apsrev4-2}

%


\end{document}